\documentclass[aps,pra,twocolumn,groupedaddress]{revtex4-1}
\usepackage{epsfig}
\usepackage{graphicx}% Include figure files
\usepackage{dcolumn}% Align table columns on decimal point
\usepackage{bm}% bold math
\usepackage{amsthm}
\usepackage{amsfonts}
\usepackage{amscd}
\usepackage{amsmath}    % need for subequations
\usepackage{enumerate}

\newcommand{\ket}[1]{\mbox{$ | #1 \rangle $}}

\begin{document}

%\preprint{}

\title{Measurement device independent quantum key distribution}



\author{Hoi-Kwong Lo$^{1}$, Marcos Curty$^{2}$, and Bing Qi$^{1}$}
%\thanks{}
\affiliation{$^1$Center for Quantum Information and Quantum Control,
Dept. of Electrical \& Computer Engineering and Dept. of Physics, University of Toronto, Toronto, Ontario, M5S 3G4, Canada \\
$^2$Escuela de Ingenier{\'i}a de Telecomunicaci\'on, Dept. of Signal Theory and Communications,
University of Vigo, Vigo, Pontevedra, 36310, Spain
%\\
%$^3$ Center for Quantum Information and Quantum Control, Department of Physics and Department of
%Electrical \& Computer Engineering, University of Toronto, M5S 3G4 Toronto, Ontario, Canada
}


\date{\today}

\begin{abstract}
How to remove detector side channel attacks has been a notoriously hard problem in quantum cryptography. Here, we propose a simple solution to this problem---{\it measurement} device independent quantum key distribution. It not only removes all detector side channels, but also doubles the secure distance with conventional lasers. Our proposal can be implemented with standard optical components with low detection efficiency and highly lossy channels. In contrast to the previous solution of full device independent QKD, the realization of our idea does not require detectors of near unity detection efficiency in combination with a qubit amplifier (based on teleportation) or a quantum non-demolition measurement of the number of photons in a pulse. Furthermore, its key generation rate is many orders of magnitude higher than that based on full device independent QKD. The results show that long-distance quantum cryptography over say 200km will remain secure even with seriously flawed detectors.
\end{abstract}

%\keywords{}

\maketitle

%\section{Introduction}\label{intro}

Quantum key distribution (QKD) allows two parties (typically called Alice and Bob)
 to generate a common string of secret bits, called
 a secret key, in the presence of an eavesdropper, Eve \cite{qkd1}.
% Such a
 This key can
 %subsequently
 be used for
 tasks such as secure communication and authentication.
% In 1984, Bennett and Brassard published the first QKD protocol, known as BB84 \cite{BB84}.
%% In 1991, Ekert independently proposed a QKD scheme based on entanglement \cite{ekert}.
% Since then, research interest in QKD has grown exponentially and
%% Experimental QKD has been performed over long distances \cite{long_qkd1,long_qkd2,long_qkd3,long_qkd4,long_qkd5}.
% commercial QKD systems are already on the market \cite{commercial}.
Unfortunately, there is a big gap between the theory and practice of QKD.
 In principle, QKD offers unconditional security guaranteed by the laws of physics \cite{lo-chau,shor,mayers}.
 However, real-life implementations of QKD
% systems
 rarely conform to the
 assumptions in idealized models used in security proofs.
% In particular, standard security proofs often assume that Alice and Bob have
% almost perfect control on the state preparation and of the measurement devices.
%Recently, there has been a lot of interest in ``quantum hacking".
 Indeed, by exploiting security loopholes in
% real-life implementations,
practical realizations, especially imperfections in the detectors,
different attacks have been successfully launched against commercial QKD systems
\cite{hack1,hack3}, thus highlighting their practical vulnerabilities.

%, including, for instance, the time-shift attack \cite{hack3,hack4},
% the phase re-mapping attack \cite{hack1,hack2},
% and the blinding attack \cite{}, .
% These hacking strategies highlight the disconnect between the theory and practice of
% the security of QKD schemes.

To connect theory with practice again, several approaches have been proposed.
 The first one is the presumably hard-verifiable problem of
 trying to characterize real devices
 fully and account for all side channels.
%
% to construct better theoretical models of employed devices in a
% QKD system.
%% Indeed, issues such as detection efficiency mismatch can, in principle, be taken care of
%%% in such an approach
%% through better characterization of a measurement apparatus \cite{efficiency_mismatch}.
%Nonetheless, such devices are often complicated solid-state systems and it may be hard to characterize them
%fully and account for all side channels.
The second approach is a teleportation trick \cite{lo-chau,note_t}.
The third solution is (full) device independent QKD (DI-QKD) \cite{diqkd1}.
% It is called device independent because
% it
 This last technique
 does not require detailed knowledge of how QKD devices work and can prove
 security based on the violation of a Bell inequality.
% but demands only that
% quantum mechanics to be correct, the choice of measurement basis to be random, and that
% there are no unwanted signals leaking to Eve from Alice and Bob's laboratories.
% DI-QKD highlights the deep connection between the foundations of quantum mechanics (particularly,
% the testing of Bell inequalities) and the security of QKD.
% DI-QKD has attracted immense theoretical interest (see, e.g. \cite{qubitamp1,qubitamp2,qubitamp3} and references cited therein).
Unfortunately,
%this solution is highly impractical as
%it requires a
DI-QKD is highly impractical because it
needs near
unity detection efficiency
together with a qubit
amplifier or a quantum non-demolition (QND) measurement of the number of
photons in a pulse,
and even then generates an extremely low key rate (of order $10^{-10}$ bits per pulse)
at practical distances \cite{qubitamp1}.
%Moreover,
%such requirement of
%almost perfect detection efficiency already assumes the
%use of a qubit
%amplifier (based on teleportation) or a quantum non-demolition (QND) measurement of the number of
%photons in a pulse. Otherwise,
%even if one considers detectors with perfect efficiency,
%the losses
%in the quantum channel would render DI-QKD useless at around 5 km.
%For this reason,
%DI-QKD
%has never been implemented in practice
%and is not regarded as practical.

In this Letter we present the idea of
measurement device independent QKD (MDI-QKD)
as a simple solution to remove all (existing and yet to be discovered) detector side channels \cite{hack3}, arguably the most critical part of the implementation,
and show that it has both excellent security and performance.
% Similar to standard DI-QKD,
%% MDI-QKD
% it
% allows Alice and Bob's measurement apparatuses to function in an arbitrary manner.
% Notice that most side channel attacks occur in detectors
% \cite{hack3}. MDI-QKD removes all such (existing and yet to be discovered)
% detector side channels automatically.
 Therefore,
 it offers an immense security advantage over standard
 security proofs such as
 Inamori-L\"utkenhaus-Mayers (ILM) \cite{ilm} and
Gottesman-Lo-L\"utkenhaus-Preskill (GLLP) \cite{gllp}. Furthermore, it has the power to
 double the transmission distance that can be covered by those QKD schemes that use conventional laser diodes,
% . This is done by simply placing an untrusted Bell state analyzer in the middle of Alice and Bob. Furthermore,
% the
and its key generation rate
% of
%MDI-QKD
%is many orders of magnitude higher than that of DI-QKD and
is comparable to that of standard security
proofs with entangled pairs.
%Also, unlike DI-QKD, MDI-QKD does not require near unity detection efficiency together
%with a qubit amplifier or a QND measurement.
In contrast to
%standard
DI-QKD, in its simplest formulation MDI-QKD
 requires the additional assumption that Alice and Bob have almost perfect state preparation. However, we believe that this is only a minor drawback because Alice's and Bob's signal sources can be
 attenuated laser pulses prepared by themselves. Their states can thus be experimentally
 verified in a fully protected laboratory environment outside Eve's interference
 through random sampling.
 Moreover, as will be discussed later, imperfections in Alice's and Bob's preparation process can, in fact, be readily taken care of in a more refined formulation of the protocol.
% MDI-QKD.

A simple example of our method is as follows. Both Alice and Bob prepare phase randomized
weak coherent pulses (WCPs) in the four possible BB84
polarization states \cite{BB84}
and send them to an {\it untrusted} relay Charlie (or Eve) located in the middle, who performs a
Bell state measurement that projects the incoming signals into a Bell state \cite{note_last}.
Such measurement can be realized, for instance, using only
linear optical elements with say the setup given in Fig.~1.
%\cite{relay}.
\begin{figure}[!t]\center
\resizebox{4cm}{!}{\includegraphics{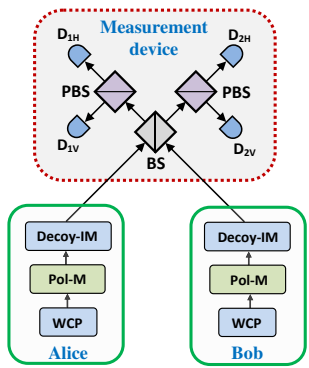}} \caption{
Basic setup of a MDI-QKD protocol. Alice and Bob prepare
phase randomized weak coherent pulses (WCPs)
%sources
%which emit laser pulses prepared
in a different
BB84 polarization state which is selected, independently and at random for each signal, by means of a
polarization modulator (Pol-M). Decoy states are generated using an intensity modulator (Decoy-IM). Inside the measurement device, signals from Alice and Bob interfere at a 50:50 beam splitter (BS) that has on each end a polarizing beam splitter (PBS) projecting the input photons into either horizontal ($H$) or 
vertical ($V$) polarization states. Four single-photon detectors
%($D_{1H}$, $D_{1V}$, $D_{2H}$ and $D_{2V}$)
are employed to detect the photons and the detection results are publicly announced. A successful Bell state measurement
corresponds to the observation of precisely two detectors (associated to orthogonal polarizations)
being triggered. A click in $D_{1H}$ and $D_{2V}$, or in $D_{1V}$ and $D_{2H}$, indicates a projection into the Bell state
$\ket{\psi^-}=1/\sqrt{2}(\ket{HV}-\ket{VH})$, while a click in $D_{1H}$ and $D_{1V}$, or in $D_{2H}$ and $D_{2V}$, reveals a projection into the Bell state
$\ket{\psi^+}=1/\sqrt{2}(\ket{HV}+\ket{VH})$.
%\cite{relay}.
Alice's and Bob's laboratories are well shielded from the eavesdropper,
while the measurement device can be untrusted.}
\end{figure}
%It consists of a beam splitter (BS) followed by two polarizing beam splitters (PBSs), one for each output arm, and four single-photon detectors, one for each output arm of the PBSs. All measurements are performed in the rectilinear basis.
(Actually, such setup only identifies two of the four Bell states. But, this is fine as any Bell state will allow a security proof to go through.) Furthermore, Alice and Bob apply decoy state techniques \cite{decoy1} to estimate the gain (i.e., the probability that the relay outputs a successful result)
and quantum bit error rate (QBER)
%(i.e., the error rate in the quantum signals)
for various input photon numbers.
%Note that decoy states can be generated experimentally by Alice and Bob
%simply through the application of intensity modulators to
%attenuate laser pulses.

%With the above intuition in mind, let us discuss our protocol in more detail.
Once the quantum communication phase is completed, Charles uses a public channel to announce
%Post-selection should be
%done by Charles' public announcement
%through
%an authenticated public channel of
the events where he has
obtained a successful outcome in the relay, including
as well his measurement result.
%(i.e., whether it is a singlet $\ket{\psi^-}$ or a triplet $\ket{\psi^+}$).
Alice and Bob keep the data that correspond to these instances and discard the rest.
Moreover,
as in BB84, they post-select the events where they use the same basis in their transmission
by means of an authenticated public channel.
Finally, to guarantee that their bit strings are correctly correlated,
either Alice or Bob
has to apply a bit flip to her/his data, except for the cases where both of
them select the diagonal basis and Charles obtains a successful measurement outcome corresponding
to a triplet state. This is illustrated in Table~1.
\begin{table}[htb]
  \centering
  \begin{tabular}{ccccc}
  \hline\hline
  Alice \& Bob & Relay output $\ket{\psi^-}$ & Relay output $\ket{\psi^+}$ & \quad  \\
  \hline
  {\rm Rectilinear basis} & {\rm Bit flip} & {\rm Bit flip} \\%& {\it Key generation basis}  \\
  {\rm Diagonal basis} & {\rm Bit flip} & - \\ %& {\it Testing basis}  \\
  \hline\hline
\end{tabular}
\caption{
Alice and Bob post-select the events where the relay outputs a successful result
and they use the same basis in their transmission. Moreover,
%to guarantee that their bit strings are correctly correlated,
either Alice or Bob flips her/his bits except for the cases where both of
them select the diagonal basis and the relay outputs a triplet.
%The rectilinear basis is used as the key generation basis, while
%the diagonal basis is used for testing only.
  }\label{table1}
\end{table}

Let us now evaluate the performance of the protocol above in detail.
The proof of its unconditional security is shown in Appendix~\ref{ap_sec}.
For simplicity, we consider a refined data analysis where Alice and Bob
evaluate the
%transmission
data sent in two bases {\it separately} \cite{efficient}.
In particular, we use the rectilinear basis as the key generation basis, while
the diagonal basis is used for testing only.
A piece of notation:
Let us denote by $Q^{n,m}_{\rm rect}$, $Q^{n,m}_{\rm diag}$,
$e^{n,m}_{\rm rect}$ and $e^{n,m}_{\rm diag}$, the gain and QBER,
respectively, of the signal states sent by Alice and Bob, where
$n$ and $m$ denote the number of photons sent by the legitimate users,
and rect/diag represents their basis choice.

{\it (A) Rectilinear basis:}
%In this case,
%it can be shown that
An error corresponds to a successful relay output when
both Alice and Bob prepare the same polarization state (i.e., their results should be
anti-correlated before they apply a bit flip). Assuming for the moment ideal optical elements and
detectors, and no misalignment,
we have that whenever Alice and Bob send, respectively, $n$ and $m$ photons prepared in
the same polarization state
the relay will never output a successful result.
%(i.e., it will
%never detect two orthogonal photons in the rectilinear
%basis).
%Also, when
%Alice sends a two-photon state and Bob emits
%a vacuum state the relay outputs an unsuccessful result.
We obtain then that $e^{n,m}_{\rm rect}$
%$e^{1,1}_{\rm rect} =e^{2,0}_{\rm rect} = e^{0, 2}_{\rm rect}$
is zero for all $n,m$. This means that no error correction is needed for the sifted key.
This is remarkable because it implies that the usage of
WCP sources (rather than single-photon sources) does not substantially lower
the key generation rate of the QKD protocol (in the error correction part).

{\it (B) Diagonal basis:}
To work out the amount of privacy amplification needed we examine the diagonal basis.
An error corresponds to a projection into the singlet state
%$\ket{\psi^-}$
given that Alice and Bob prepared the same polarization state, or into the triplet state
%$\ket{\psi^+}$
 when they prepare orthogonal polarizations.
Assuming again the
ideal scenario discussed in the previous paragraph, we find that $e^{1,1}_{\rm diag}=0$.
(This is because when two identical single-photons
%wave packets
%of the same polarization
enter a 50:50 BS
%and the temporal overlap of the photons on the BS is perfect,
the Hong-Ou-Mandel (HOM) effect \cite{HOM} ensures that
both photons will always exit the BS {\it together} in the same output mode.
%So, they will never exit in {\it different} output arms
%and be regarded as a singlet by the quantum relay.
Also, if the two photons
are prepared in orthogonal polarizations and they exit the 50:50 BS
in the same output arm, both photons will always
reach the same detector within the relay.)
The fact that $e^{1,1}_{\rm diag}$ is zero is again remarkable as it means
that the usage of WCP sources does not substantially lower the key generation
(in also the privacy amplification part).

{\it (C) Key generation rate:}
%If we ignore contributions from 3 or more photons,
%(i.e., $n+ m \geq 3$),
In the ideal scenario described above the key generation rate will be simply given by
%\begin{equation}
$R=Q^{1,1}_{\rm rect}$
%\end{equation}
in the asymptotic limit of an infinitely long key. 
On the other hand, if we take imperfections such as basis misalignment
and dark counts into account, the key generation rate
in a realistic setup
will be given by \cite{gllp,efficient,koashi}
\begin{equation}\label{keyrate}
R = Q^{1,1}_{\rm rect} [ 1 -
H(e^{1,1}_{\rm diag}) ] - Q_{\rm rect} f(E_{\rm rect}) H(E_{\rm rect}),
\end{equation}
where $Q_{\rm rect}$ and $E_{\rm rect}$ denote, respectively, the gain and QBER in the rectilinear basis
(i.e., $Q_{\rm rect}=\sum_{n,m} Q^{n,m}_{\rm rect}$, and
$E_{\rm rect}= \sum_{n,m} Q^{n,m}_{\rm rect}e^{n,m}_{\rm rect}/Q_{\rm rect}$),
$f(E_{\rm rect}) >1$ is an inefficiency function for the error correction process,
and
$H(x)=-x\log_2{(x)}-(1-x)\log_2{(1-x)}$ is the binary Shannon entropy function.

%At short distances, most successful events in the relay arise from detecting single-photon pulses
%emitted by Alice and Bob, whose QBER can be kept small by a proper alignment of the system. So,
%we expect that $e_{1,1; diagonal}$ and $ E_{rectilinear}$ to be pretty small in the low losses regime \cite{note}.

There are few loose ends that need to be tightened up.
First, we have implicitly assumed that the decoy state method can be used to
estimate the gain $Q^{1,1}_{\rm rect}$ and the
QBER $e^{1,1}_{\rm diag}$.
Second, 
%we need to take the contributions from  3 or more photons (i.e., $n+ m \geq 3$) into
%consideration. Third, 
we need to evaluate the secret key rate given by Eq.~(\ref{keyrate}) for a realistic setup. 
Let us tighten up these loose ends here.
Indeed, it can be shown that
the technique to estimate the relevant parameters in the key rate formula is equivalent to that used in standard decoy-state QKD systems
(see Appendix~\ref{ap_est} for details).
For simulation purposes,
we consider inefficient and noisy threshold detectors and  
employ
%the following
experimental parameters
%:
%the loss
%coefficient of the quantum channel is $0.2$ dB/km, the intrinsic
%error rate due to misalignment and instability of the optical system is $1.5\%$, the detection efficiency of the quantum relay
%(i.e.,
%the transmittance of its optical
%components together with
%the efficiency of its detectors)
%is $14.5\%$,
%and the background count rate
%is $6.02\times10^{-6}$.
%These parameters are
from \cite{ursin} with the exception that \cite{ursin} considered
a free-space channel whereas here we consider a fiber-based channel with a loss of $0.2$ dB/km.
Moreover, for simplicity, 
we
assume that all detectors are equal (i.e., they have the same dark count rate and detection efficiency),
and their dark counts are, to a good approximation, independent of the incoming signals. 
Furthermore, we use an error correction protocol with inefficiency function
$f(E_{\rm rect})= 1.16$ \cite{notes_f}.
The resulting
lower bound on the secret key rate is illustrated in Fig.~2.
\begin{figure}[!t]\center
\resizebox{7,4cm}{!}{\includegraphics{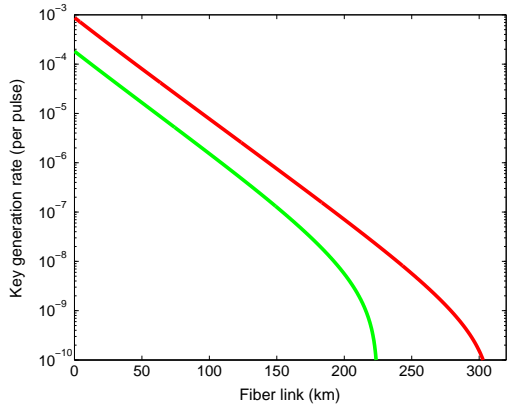}} \caption{
Lower bound on the secret key rate $R$ given by Eq.~(\ref{keyrate}) in
logarithmic scale for the MDI-QKD setup with WCPs illustrated in Fig.~1 (green curve).
For
simulation purposes, we consider the following experimental
parameters \cite{ursin}:
the loss
coefficient of the channel is $0.2$ dB/km, the intrinsic
error rate due to misalignment and instability of the optical system is $1.5\%$, the detection efficiency of the relay
(i.e.,
the transmittance of its optical
components together with
the efficiency of its detectors)
is $14.5\%$,
and the background count rate
is $6.02\times10^{-6}$. 
%We further assume an efficiency for the error correction protocol of $f(E_{\rm rect})= 1.16$.
(For simplicity, we consider a simplified model of misalignment by putting
a unitary rotation in
one of the input arms of the 50:50 BS and also a unitary rotation in
one of its output arms. The total misalignment value is $1.5\%$.
That is, we assume a misalignment of $0.75\%$ in each rotation.)
In comparison, the red curve represents a lower bound on $R$ for an entanglement-based QKD protocol with a parametric down conversion (PDC)
source
situated in the middle between Alice and Bob \cite{ma}.
In the red curve, we have assumed that an optimal brightness of a PDC source
is employed. However, in practice, the brightness of a PDC source is limited
by technology.
Therefore, the key rate of an entanglement-based QKD protocol will be
much lower than what is shown in the red curve.
This makes our new proposal even more favorable than the comparison that
is presented in the current Figure.}
\end{figure}
Our calculations and simulation results demonstrate that the key 
%generation 
rate
is highly comparable to a security proof \cite{ma} for entanglement-based QKD protocols.
Our scheme can tolerate a high optical loss of more than 40dB (i.e., 200km of optical fibers) when
a relay is placed in the middle of Alice and Bob. That is, one can essentially
double the transmission distance over a setup where the Bell measurement apparatus is on Alice's side
or a setup using a standard decoy-state BB84 protocol \cite{dark}.

%We remark that with our protocol one can essentially double the transmission distance
%of a QKD system based on WCPs over that of a
%standard decoy-state BB84 scheme.
%Indeed,
%similar to \cite{ma}, by placing the Bell measurement apparatus
%in the middle between Alice and Bob, with our scheme,
%Our work shows clearly that it is feasible for QKD with WCPs to
%achieve similar long-distance transmission than what was previously
%thought \cite{ma,long_qkd4} to be possible with only entangled states.
%Furthermore, in contrast to both the protocol of \cite{ma} and a standard decoy-state
%BB84 protocol, our new scheme
%has a tremendous advantage of being immune to all detector side channel attacks.

To experimentally implement the MDI-QKD protocol proposed,
%described above,
there are a few practical issues
that have to be addressed.
Among them, the most important one is probably how to
generate indistinguishable photons from two independent laser
sources and observe stable HOM interference
\cite{HOM}. Note that the physics behind this protocol is based on the
photon bunching effect of two indistinguishable photons at a
50:50 BS. We performed a simple proof of principle experiment to
show that a high-visibility HOM interference between two independent
off-the-shelf lasers is actually feasible (see details in Appendix~\ref{ap_exp}).
The results are shown in
Fig.~3. 
%For comparison purposes, this figure also shows the theoretical results
%that arise from the model in \cite{Rarity05}.  
The consistency
between experimental and theoretical results confirms that a
high-visibility HOM dip can be obtained even with two independent lasers.
%, thus
%demonstrating the feasibility of our QKD protocol.
\begin{figure}[!t]\center
\resizebox{8,05cm}{!}{\includegraphics{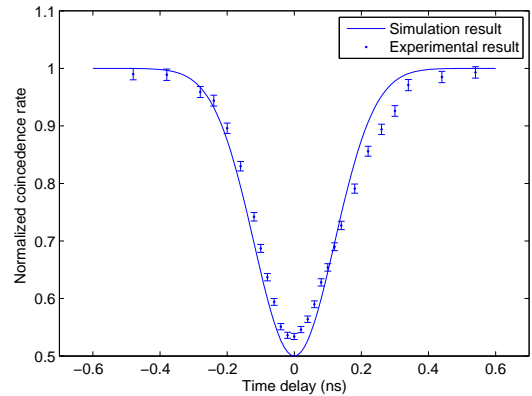}} \caption{
Hong-Ou-Mandel interference between two phase randomized WCPs. 
%The pulse shape is approximately Gaussian with a
%pulse width (FWHM) around 200ps. 
The average photon number is $0.1$
per pulse. The coincidence rate is recorded at different time
delays. 
%As a comparison, we also show the simulation results with
%perfect 200ps Gaussian pulses. 
The error bars show the statistical
fluctuation ($\pm$ one standard deviation) due to finite data size.}
\end{figure}

The idea of MDI-QKD can be generalized much further.
First of all, it also applies to the case where Alice and Bob use entangled
photon pairs
%(rather than WCPs)
as sources.
Second, it works even when Alice and Bob's preparation processes
are imperfect. Indeed, basis-dependence that originates from
the imperfection in Alice and Bob's preparation processes can be readily taken care of
by using a quantum coin idea \cite{gllp,koashi} to
quantify the amount of basis-dependent flaw \cite{tamaki}. Third, notice that in practical
applications only a finite number of decoy states will be needed. This is
similar to standard finite decoy state QKD protocols \cite{practical_decoy} that have been widely employed in experiments \cite{long_qkd5}.
Fourth, MDI-QKD works even without a refined data analysis.
%Fifth, MDI-QKD works also for a simpler experimental setup with a single BS followed by two single-photon detectors, one at each arm.
Fifth, it works also for other QKD protocols including the six-state protocol \cite{six}.
These subjects will be discussed further in future publications.

In summary, we have proposed the idea of measurement device independent QKD (MDI-QKD).
Compared to standard security proofs, it has a key advantage of removing all detector side channels,
and it can double the transmission distance covered with conventional QKD schemes using WCPs.
Moreover, it has a rather high key generation rate which is comparable to
that of standard security proofs. Indeed, its key generation rate is orders of magnitude
higher than the previous approach of full device independent QKD.
Our idea can be implemented with standard threshold detectors with low detection efficiency
and highly lossy channels.
%In contrast, full device independent QKD
%requires
%detectors of near unity detection efficiency in combination with a
%qubit amplifier or QND measurements.
In view of its excellent security, performance and simple implementation,
we believe MDI-QKD is a big step forward in bridging the gap between the theory and
 practice of QKD, and
we expect to be widely employed in
practical QKD systems in the future.

{\it Notes Added}: After the posting of our paper on public preprint servers,
another paper by Braunstein and Pirandola \cite{braunstein}
has been posted on the preprint servers \cite{compare}.

Useful conversations with C. H. Bennett, H. F. Chau, C.-H. F. Fung, P. Kwiat, X. Ma, K. Tamaki and Y. Zhao are gratefully acknowledged. We thank N. L\"utkenhaus for discussions about Inamori's security proof, C. Weedbrook for comments on the presentation of the paper, and Z. Liao and Z. Tang for helping us in some parts of the experiment.
We thank NSERC, the Canada Research Chair Program, QuantumWorks, CIFAR and
Xunta de Galicia for financial support.

\appendix

\section{Security analysis}\label{ap_sec}

Here, we show that our protocol for measurement device independent quantum key distribution (MDI-QKD)
is unconditionally secure.
% In one word,
% our key idea is to
Our
 security proof
% based on
 is inspired by that of a time reversed EPR-based QKD protocol \cite{inamori} and the decoy state method
 \cite{decoy1}. It can be
% implemented with
applied to practical phase randomized weak coherent pulses (WCPs)
generated by a laser \cite{note_m}.

%Why is our protocol secure?
%The intuition of its security is as follows.

In the protocol, each of Alice and Bob uses decoy states
and WCPs to prepare the four BB84 states and
sends their state to Charles. Charles combines the two signals sent by Alice and Bob and
performs a Bell measurement. Afterwards, he announces in a public channel whether he
has received a Bell state and which specific Bell state he has received,
say, for instance, a singlet state $  {1/\sqrt{2}} (\ket{HV} - \ket{VH})$.
Alice and Bob post-select only those transmission events where Charles has received
some specific Bell state.

Using decoy states, Alice and Bob can now
obtain the gain and QBER of those events where both of them send to Charlie {\it single}-photon states. As in GLLP \cite{gllp}, let us consider a {\it virtual} qubit idea.
Instead of preparing a single-photon BB84 state, Alice prepares
its purification. That is to say that
one could imagine that Alice actually has a virtual qubit on her side
and she prepares her state by first preparing an entangled state
of the combined system of her virtual qubit and the qubit that she
is sending out in say a singlet state. She subsequently measures
her virtual qubit, thus preparing a BB84 state. (Similarly, Bob uses a
virtual qubit to help him prepare a single-photon BB84 state.)

Now, in principle, Alice could as well keep her virtual qubit in her quantum
memory and delay her measurement on it.
Only after Charles has announced that he has obtained a successful
outcome (say a singlet), will Alice perform a measurement on her virtual
qubit to decide on which state she is sending to Bob.

In such virtual qubits setting, the protocol is directly equivalent to an
entanglement based protocol \cite{lo-chau,shor,intuition}. 
Alice and Bob share a pair of qubits in their quantum memories
and they simply compute the QBER on their virtual qubits in the $XX$ and $ZZ$ bases.

Furthermore, with the above virtual qubit picture in mind,
one sees that what Alice and Bob actually send out is unimportant
for security proofs as long as their single-photon signals are basis-independent.
In the event that there are some basis-dependent flaws in their
preparation, Alice and Bob can take care of them by using a quantum coin
idea \cite{gllp,tamaki2}. (Notice that GLLP \cite{gllp}
only considers imperfect state preparation by Alice.
Here we are interested in simultaneous imperfections in both Alice and Bob.
Nonetheless, with the above virtual qubit formulation, such simultaneous imperfections
can still be taken care of by the quantum coin idea. One only needs to
consider the fidelity between the combined states sent out by Alice
and Bob $\rho^X_{AB}$ and
$\rho^Z_{AB}$, where Alice and Bob both use either the $X$ or $Z$ basis.
For a more detailed discussion, see, for example, \cite{tamaki2}.)

%Another important advantage of
Furthermore, our protocol
%over existing schemes
%is that it
can tolerate very high channel loss and very low success probability of a Bell measurement without compromising its security. This is because
Alice and Bob can post-select only those
successful events for a Bell-state measurement by Charles for consideration.
Note that Alice and Bob could have implemented their protocol with virtual qubits in their quantum memories and could have waited until Charles announces which Bell-state measurement
results are successful. Indeed, in security proofs
such as \cite{lo-chau} and \cite{shor}, losses do not affect security.

\section{Estimation procedure}\label{ap_est}

Here we show that the decoy state method \cite{decoy1} applied to MDI-QKD
allows Alice and Bob to estimate the relevant parameters to evaluate the secret key rate formula in the
asymptotic regime. In particular, they can obtain the gain $Q^{1,1}_{\rm rect}$ and the
QBER $e^{1,1}_{\rm diag}$.

Our starting point is the standard decoy state technique applied to conventional QKD.
We assume it permits Alice and Bob to estimate the yield $Y_n$ and error rate $e_n$
of an $n$-photon signal for all $n$ \cite{decoy1}. That is, the set of linear equations
\begin{eqnarray}%\label{decoy}
Q^i=\sum_{n=0}^\infty e^{-\mu_i} \frac{\mu_i^n}{n!} Y_n, \label{decoy1} \\
Q^iE^i=\sum_{n=0}^\infty e^{-\mu_i} \frac{\mu_i^n}{n!} Y_ne_n, \label{decoy2}
\end{eqnarray}
with the index $i$ denoting the different decoy settings,
can be solved and Alice and Bob can obtain the parameters $Y_n$ and $e_n$ for all $n$.

In MDI-QKD we have the following set of linear equations:
\begin{eqnarray}
Q_{\rm rect}^{i,j}&=&\sum_{n,m=0}^\infty e^{-\mu_i} \frac{\mu_i^n}{n!}e^{-\mu_j} \frac{\mu_j^m}{m!} Y^{n,m}_{\rm rect},  \\
Q_{\rm diag}^{i,j}&=&\sum_{n,m=0}^\infty e^{-\mu_i} \frac{\mu_i^n}{n!}e^{-\mu_j} \frac{\mu_j^m}{m!} Y^{n,m}_{\rm diag},
\end{eqnarray}
and
\begin{eqnarray}
Q_{\rm rect}^{i,j}E_{\rm rect}^{i,j}&=&\sum_{n,m=0}^\infty e^{-\mu_i} \frac{\mu_i^n}{n!}e^{-\mu_j} \frac{\mu_j^m}{m!} Y^{n,m}_{\rm rect} e^{n,m}_{\rm rect}, \nonumber \\
Q_{\rm diag}^{i,j}E_{\rm diag}^{i,j}&=&\sum_{n,m=0}^\infty e^{-\mu_i} \frac{\mu_i^n}{n!}e^{-\mu_j} \frac{\mu_j^m}{m!} Y^{n,m}_{\rm diag}e^{n,m}_{\rm diag}, \nonumber
\end{eqnarray}
where the indexes $i$ and $j$ represent, respectively, the different decoy settings used by Alice and Bob.

Let us begin with the gain $Q_{\rm rect}^{i,j}$. This quantity can be written as
\begin{equation}\label{v1}
Q_{\rm rect}^{i,j}=\sum_{n=0}^\infty e^{-\mu_i} \frac{\mu_i^n}{n!}Y_{n; \rm rect}^j,
\end{equation}
where
\begin{equation}\label{v2}
Y_{n; \rm rect}^j=\sum_{m=0}^\infty e^{-\mu_j} \frac{\mu_j^m}{m!} Y^{n,m}_{\rm rect}.
\end{equation}
For $j$ fixed, varying $i$ we have that Eq.~(\ref{v1}) is equivalent to
Eq.~(\ref{decoy1}). This means that Alice and Bob can estimate the parameters
$Y_{n; \rm rect}^j$. Once the yields $Y_{n; \rm rect}^j$ are obtained  for all $j$,
we have that
Eq.~(\ref{v2}) is again equivalent to
Eq.~(\ref{decoy1}) and the legitimate users can estimate the parameters $Y^{n,m}_{\rm rect}$.

Similarly, it can be shown that Alice and Bob can estimate $Y^{n,m}_{\rm diag}$.

Let us now focus on the QBER $E_{\rm rect}^{i,j}$. It can be written as
\begin{equation}\label{v3}
Q_{\rm rect}^{i,j}E_{\rm rect}^{i,j}=\sum_{n=0}^\infty e^{-\mu_i} \frac{\mu_i^n}{n!} W_{n; \rm rect}^j,
\end{equation}
where
\begin{equation}\label{v4}
W_{n; \rm rect}^j=\sum_{m=0}^\infty e^{-\mu_j} \frac{\mu_j^m}{m!} Y^{n,m}_{\rm rect} e^{n,m}_{\rm rect}.
\end{equation}
Again, for $j$ fixed, varying $i$ we have that Eq.~(\ref{v3}) is equivalent to
Eq.~(\ref{decoy1}), so Alice and Bob can estimate the parameters
$W_{n; \rm rect}^j$. Once these quantities are obtained for all $j$, we have that
Eq.~(\ref{v4}) is equivalent to
Eq.~(\ref{decoy2}) with the yields $Y^{n,m}_{\rm rect}$ already known. This means that Alice and Bob
can estimate
$e^{n,m}_{\rm rect}$.

Similarly, it can be shown that Alice and Bob can estimate $e^{n,m}_{\rm diag}$.

We have demonstrated that
Alice and Bob can estimate the parameters $Y^{n,m}_{\rm rect}$, $Y^{n,m}_{\rm diag}$, $e^{n,m}_{\rm rect}$,
and $e^{n,m}_{\rm diag}$ for all $n$ and $m$ in the asymptotic case; in particular, they can obtain the relevant quantities
$Y^{1,1}_{\rm rect}$ and $e^{1,1}_{\rm diag}$. Now, the gain $Q^{1,1}_{\rm rect}$
is given by
\begin{equation}
Q^{1,1}_{\rm rect}=\mu_A\mu_Be^{-(\mu_A+\mu_B)}Y^{1,1}_{\rm rect},
\end{equation}
where $\mu_A$ and $\mu_B$ denote, respectively, the mean photon number of Alice and Bob's signals.

In practical
applications only a finite number of decoy states will be needed. This is
similar to standard finite decoy state QKD protocols \cite{practical_decoy,decoy_experiment}.

\section{Experimental setup of proof of concept experiment}\label{ap_exp}

For our idea of MDI-QKD to work in practice, it is important to be able to obtain interference
between signals generated by two {\it independent} lasers (on Alice's and Bob's sides) \cite{inter}.
Notice that this requirement is at the heart of many quantum information applications such as quantum repeaters, teleportation, qubit amplifiers for device independent QKD, etc.
Many investigations and a lot of progress have been made on this subject in recent years
(see, {\it e.g.}, \cite{gisin_mod} and references therein).

To show that photons from two independent laser sources can be made
indistinguishable and interfere with each other, we have performed a
simple proof of concept experiment. The experimental setup is shown
in Fig.~1. 
\begin{figure}[!t]\center
\resizebox{8cm}{!}{\includegraphics{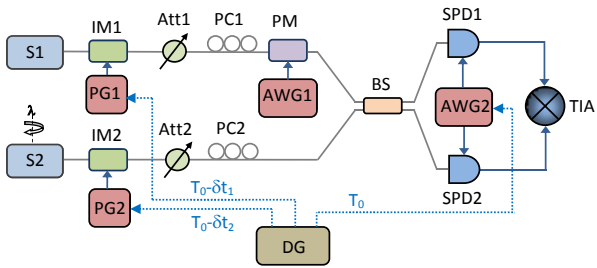}} \caption{
Experimental setup. S1: 1550nm cw fiber laser (NP photonics); S2:
wavelength tunable cw laser (Agilent); IM1-IM2: optical intensity
modulators; Att1-Att2: variable optical attenuators; PC1-PC2:
polarization controllers; PM: optical phase modulator; BS: 50:50
fiber beam splitter; SPD1-SPD2: single-photon detectors; TIA: Time
Interval Analyzer (Picoquant); PG1-PG2: electrical pulse generators
(Avtech); AWG1-AWG2: arbitrary waveform generators (Agilent); DG:
delay generator (Stanford research systems).}
\end{figure}
S1 denotes a 1550nm cw fiber laser (NP photonics) and
S2 is a wavelength tunable cw laser (Agilent). Both lasers have
linewidths less than 1MHz and there is no optical or electrical
connection between them. The central wavelength of S2 can be
adjusted at steps of 0.1pm (which corresponds to 13MHz at 1550nm).
Two optical intensity modulators (IM1 and IM2) are employed
to generate narrow laser pulses from the above two cw lasers. Each
of these two intensity modulators is driven by an electrical pulse
generator (Avtech, PG1 and PG2), which is triggered by a
4-channel delay generator (DG). The time delay between the
two laser pulse trains can be adjusted by the delay generator at ps
resolution. Two variable optical attenuators (Att1 and Att2) are used to determine the average photon number per laser
pulse. An optical phase modulator (PM) is used to scan the
phase of one laser pulse train. The two laser pulse trains interfere
at a symmetric 2x2 fiber coupler (BS), and the
interference signals are detected by two InGaAs single-photon
detectors (SPD, ID Quantique). A time interval analyzer (TIA,
Picoquant) is employed to perform a coincident measurement. The whole
system is built upon single-mode telecom fiber based components,
thus the spatial modes of the two laser beams are identical. By
using two polarization controllers (PC1 and PC2), we can
ensure that the two laser pulse trains are in the same polarization
state when they interfere at the BS.
In this experiment, Alice and Bob are on the same
optical table and the fiber length from Alice/Bob to the measurement
device is only a few meters. 

The pulse shapes of the laser pulses that output from the two intensity
modulators have been carefully matched by adjusting the two
electrical pulse generators. This shape is approximately
Gaussian with a pulse width (FWHM) around 200ps. The corresponding
spectral bandwidth is about 5GHz, which is significantly larger than
the central frequency mismatch (below 30MHz during the whole
experiment). Thus the photons from the two lasers are indistinguishable
in the spectral domain.

The MDI-QKD protocol proposed encodes quantum information in phase
randomized WCPs. This means that the phase should be randomly
changed from pulse to pulse. This can be
achieved by using an optical phase modulator after each laser to
introduce a random phase shift from pulse to pulse, as we have
demonstrated previously \cite{Zhao07}. Here, for simplicity, the
phase shift is scanned periodically in the range of $[0, 2\pi]$ at
900Hz. In our experiment, 
the laser pulse repetition rate is 500KHz and
one measurement takes about 500s, thus the
phase is equivalently randomized.

We measure the Hong-Ou-Mandel (HOM) interference at an average photon
number of 0.1 per pulse. During the experiment, at each time delay $\delta_{t_2}-\delta_{t_1}$,
we record 
the detection probability of SPD1 ($P_1$), the detection probability
of SPD2 ($P_2$), and the probability of having a simultaneous ``click" in both SPDs within a coincidence window of 2ns ($P_C$).
The normalized coincidence rate is calculated as
$C=P_C/(P_1P_2)$. The experimental results
are shown in Fig.~3 of the Main Text of our paper. For comparison purposes, 
this figure also includes the theoretical 
results obtained using perfect 200ps
Gaussian pulses. The error bars arise from
statistical fluctuations ($\pm$ one standard deviation) due to finite
data size. The measured HOM dip is $0.534\pm0.005$. This result 
confirms that a high-visibility HOM dip can be obtained with independent 
lasers, thus demonstrating the feasibility of our QKD protocol.
%
%visibility is
%good enough for generating secure key in MDI-QKD.
%\begin{figure}[!t]\center
%\resizebox{7.5cm}{!}{\includegraphics{HOM.eps}} \caption{The
%Hong-Ou-Mandel interference between two phase randomized weak
%coherent pulses. The pulse shape is approximately Gaussian with a
%pulse width (FWHM) around 200ps. The average photon number is 0.1
%per pulse. The coincidence rate is recorded at different time
%delays. As a comparison, we also show the simulation results with
%perfect 200ps Gaussian pulses. The error bars show the statistical
%fluctuation (+/- one standard deviation) due to finite data size.}
%\end{figure}

The major error sources in the experiment seem to be the
pulse shape mismatch and time jitter. Limited by the equipment
available in our lab, PG1 and PG2 are actually two different pulse generators which
has limited control over pulse shapes. Other sources of errors are
the finite extinction ratio of the intensity modulator, the frequency
mismatch between the two lasers, the polarization mismatch, the asymmetry of
the BS, the error in determining the average photon number,
the dark counts of the SPDs, and statistical fluctuations due to finite data size.
%
%etc. As a future research project, we will quantify the contribution
%of each error source and improve the system design.

%In our proof-of-principle experiment, Alice and Bob are on the same
%optical table and the fiber length from Alice/Bob to the measurement
%device is only a few meters. Notice that typically the signal
%strength in a conventional decoy state technique is around $0.5$
%photon per pulse. So, the signal strength that we are using is not
%entirely unreasonable (at least for short-distance implementation
%when the loss of the channel is limited). Therefore, our proof of
%principle experiment demonstrates that at least for a short-distance
%implementation, it is feasible to see HOM-dip with independent
%lasers.

%%%%%%%%%%


\begin{thebibliography}{99}
\bibitem{note_t}
Instead of receiving quantum
 signals from Alice through a quantum channel directly, Bob
 %sets up a teleportation gateway inside his own
 %secured laboratory. He
 %prepares entangled pairs and
 teleports any incoming signal from
 outside to himself.
% Note that teleportation produces only classical bits. Consequently, Bob receives only
% classical data (rather than quantum signals) from outside.
% Furthermore, teleportation provides an exact counting on the dimension of the received
% Hilbert space.
 Provided that Bob has perfect control on the state preparation,
% prepared his entangled pairs correctly,
 he can remove all side-channels through teleportation.
 While this proposal is technologically feasible, a high-performance
% speed, high-efficiency, and
% high-fidelity
 implementation of teleportation is not without its own challenges \cite{teleportation}.
\bibitem{note_last} Note that a key advantage of our work is that Charles'
detection system can be arbitrarily flawed without compromising security.
\bibitem{notes_f}
In practice, the effect of the inefficiency function $f(E_{\rm rect})$ on the key rate is very small.
Notice that the value $f(E_{\rm rect})= 1.16$ that we use here is rather conservative. One can
achieve better results with, for instance, good error correcting codes.
For a given distance, we optimize the intensity of Alice's and Bob's lasers numerically to maximize the key rate. As expected,
such intensities are on the order of one.
\bibitem{compare}
Unlike our paper, the manuscript \cite{braunstein} does not consider decoy
state or
weak coherent pulses. So, it is unclear what key rate it will give.
Moreover, the authors of \cite{braunstein} appear to be unaware
of previous works by Biham \cite{biham} and by Inamori \cite{inamori}. So,
\cite{braunstein} does not discuss about its difference from those
important previous papers.
\bibitem{intuition}
The intuition of Shor-Preskill's proof \cite{shor} is the following.
First of all, note that if Alice and Bob share $k$ pairs of singlets (i.e., Alice
has one half of each pair and Bob has the other half.),
then they can generate a $k$-bit key that is secure from any
eavesdropper, Eve, by measuring their $k$ pairs
in the same basis. The key point is that entanglement is monogamous and Eve cannot clone such
quantum correlation. Now, suppose
two distant parties, Alice and Bob, share a large number, say $N$ of pairs of qubits
that are some noisy version of $N$ singlets. They may apply
some random sampling procedure to
estimate the average state of the qubit pairs that they share,
i.e., they can randomly choose a smaller number, say $m$, of their pairs for testing.
Based on the estimated average state, Alice and Bob can then
apply a procedure called entanglement distillation protocol
(EDP) to distill a smaller number, say $k$, of qubit pairs that are very close in
fidelity to a $k$-singlet state.
Notice that a general EDP requires a quantum computer \cite{lo-chau}.
Fortunately, Shor-Preskill showed how one can do a special class of
EDPs without using a quantum computer.
The class of EDPs used by Shor-Preskill involve the separate error correction of
two types of errors, namely bit-flip and phase errors.
\end{thebibliography}
\bibliographystyle{apsrev}

\end{document}